\documentstyle[preprint,aps]{revtex}
\begin{document}
\draft
\preprint{UTPT-97-13}
\title{On the Relativistic Collapse of Dense Star Clusters}
\author{J. W. Moffat}
\address{Department of Physics, University of Toronto,
Toronto, Ontario M5S 1A7, Canada}

\date{\today}
\maketitle

\begin{abstract}%
We investigate the fate of a relativistic star cluster with a dense core
which is undergoing a gravothermal catastrophe and is far from
thermodynamic equilibrium. Nonlinear cooperative contributions 
are included in the standard transport equations for the last stage of evolution of a
highly dense core of stellar remnants. We find that the core redshift does not necessarily increase
without limit as the core becomes increasingly dense, preventing collapse to a black hole. In
particular, the redshift can remain less than the critical value for relativistic collapse, resulting in
a stable, massive dark core with a Newtonian mantle and halo.
\end{abstract}
\vskip 0.2 true in
{\it Subject headings}: black hole physics - galaxies: nuclei - gravitation

\pacs{ }

\section{Introduction}

According to numerical integrations of the full Einstein equations for the 
dynamical evolution of an arbitrary spherical, collisionless system in general 
relativity (Shapiro and Teukolsky 1985a,b,c, 1986), star clusters become 
relativistically unstable at sufficiently high central redshift, $z_c > 0.5$,
confirming the original speculation by Zel'dovitch and 
Podurets (1965) and perturbative calculations  performed by Ipser and Thorne (1968),
Ipser (1969,1970,1980), Bisnovatyi-Kogan and Thorne (1970), Fackerell (1970) and Fackerell et
al., (1969). However, more recent work by Rasio, Shapiro and Teukolsky (1988,1989),
Bisnovatyi-Kogan, et al., (1993) and Merafina and Ruffini (1995), demonstrated that a family of
cluster evolution models were relativistically stable for $z_c\rightarrow\infty$. Merafina and
Ruffini proposed three possible configurations of cluster equilibrium models in terms of central
redshift versus temperature diagrams. They found that the stable relativistic equilibrium models,
corresponding to the possibility of reaching infinite central redshift, were not of astrophysical
interest, because they have ratios $M_{\rm core}/M\sim 10 ^{-24}$ and 
$R_{\rm core}/R\sim 10 ^{-25}$, where $M_{\rm core}$ and $R_{\rm core}$ denote the core
values of the cluster mass and radius, respectively. We shall accept the point of view that 
astrophysically interesting clusters, which can evolve dense relativistic cores with masses
$10^7-10^9\,M_{\odot}$, will become unstable for $z_c > 0.5$ and collapse to black holes.

The question remains: {\it do supermassive black holes form at the center of galaxies?} In a
recent article (Moffat 1997), a model of a galaxy was constructed from a relativistic massive core
and a Newtonian mantle and halo. The core consisted of stellar remnants
such as neutron stars, brown dwarfs or planets with individual masses,
$m\sim 1\,M_{\odot}$. This would not contradict the observations obtained 
by Lugger et al., (1992), which ruled out a change in mass-to-light ratio from
$\sim 2$ at 1 arsec to $\geq 20$ at $0.1$ arcsec, yielding large broadband 
color gradients. Using the high-velocity maser emission data obtained for
NGC 4258 ($M\sim 4\times 10^7\,M_{\odot}$)(Miyoshi et al., 1995; Maoz 1995), which impose
constraints on the mass distribution of the dense 
stellar cluster, the evaporation time obtained was $t_{\rm evap}\sim 10$
Gyr, a time scale big enough compared to the Hubble time.

Calculations have been performed of dense ($\geq 10^8\,
{\rm stars}\, {\rm pc}^{-3}$) galactic nuclei composed initially of 
main-sequence stars (Gold, Axford, and Ray 1965; 
Spitzer and Stone 1967; Colgate 1967; Sanders 1970; Spitzer 1971). It was 
found that the evolution is mainly determined by coalescence and stellar 
collisions, and massive stars would undergo supernova explosions that can leave 
behind compact stellar remnants. The endpoint of this evolutionary phase is
likely to be a cluster of neutron stars or other stellar 
remnants. Such a cluster could initially consist of $10^8$ compact
objects, within a region $\leq 0.1$ pc in radius and moving with a velocity
dispersion $800 - 2000\, {\rm km}\, {\rm s}^{-1}$. Two epochs 
characterize the subsequent Newtonian evolution of the cluster of compact
stars, (i) a low-redshift ($z_c < z_{\rm coll}\approx 10^{-2}$) point-mass
epoch during which the core undergoes collapse driven by the 
``gravothermal catastrophe" (Lynden-Bell and Wood 1968; Cohn 1980), (ii) a 
subsequent short, high-redhift ($z_c \geq z_{\rm coll}$) epoch dominated by
coalescence and collisions of compact stars in the cluster.

\section{Basic Nonlinear Transport Equations}

We must now investigate the evolution of the dense core following these
two epochs. Shapiro and Teukolsky (1987), Goodman and Lee (1989),
Quinlan and Shapiro (1989), Lee (1995), Quinlan (1996), 
van der Marel et al., (1997) have studied this epoch using transport equations 
for a cluster of compact stars (Lightman and Shapiro 1978). The number
of core stars decreases and, at the same time, the central velocity dispersion
and redshift increase ultimately leading to a relativistic redshift,
$z_c > z_{\rm crit}\sim 0.5$, and therefore to gravitational collapse to a black hole. 

These authors assumed that the transport equations were {\it linear}
differential equations in the time variable $t$. The arguments against dark 
clusters of compact stars rely mainly on the very high density of the
core ($\geq 10^8\,M_{\odot}\, {\rm pc}^{-3}$). Treated as a fluid, the
gravothermal catastrophe epoch {\it is expected to be far from thermodynamic
equilibrium} (Lynden-Bell and Wood 1968).

There are many examples in physics in which subsystems cooperate with each other in a 
self-organizing way (Haken 1975; Haken 1983; Bak, Tang and Wiesenfeld 1988; Bak 1996). The
behavior of the total system can show characteristic changes which can be described as a
transition from disorder to order, or a transition from one state to another.
There are numerous examples of such behavior at both the macroscopic
and quantum levels. Pronounced cooperative phenomena may occur in
physical systems far from thermodynamic equilibrium. Such physical
systems far from equilibrium can display ordered states created and 
maintained by an energy flux passing through the system. The choices of
{\it order parameters}, well-known in phase-transition theory, play an
important role. The order parameters represent the behavior of the system
on a macroscopic scale and therefore describe macroscopic variables.
Normally the order parameters satisfy simple differential equations with
respect to the time variable, for the relaxation time of order parameters is
normally much greater than those of the subsystems.  The variables of the
subsystem can be eliminated without increasing the degree of the time 
derivatives. 

We shall generalize the differential equations for the dense core given by
Lightman and Shapiro (1978) and Shapiro and Teukolsky (1985) to the following equations:
\begin{mathletters}
\begin{eqnarray}
\label{energy}
{\dot E}_c&=&-(\alpha_1/t_r)E_c-(\beta_1/t_r^3)E^3_c+(1/t_{\rm coll})E_c,\\
\label{Nequation}
{\dot N}_c&=&-(\alpha_2/t_r)N_c-(\beta_2/t^3_r)N_c^3-(1/t_{\rm coll})N_c,\\
\label{mequation}
{\dot m}_c&=&(1/t_{\rm coll})m_c,
\end{eqnarray}
\end{mathletters}%
where $N_c(t)$ is the number of core stars, ${\dot N}_c
=\partial N_c/\partial t$, $E_c(t)$ is the characteristic core binding energy
and $m_c(t)$ is the mean stellar mass. By using the virial theorem 
we get 
\begin{equation}
E_c\approx \frac{1}{2}N_cm_cv_c^2\approx \frac{1}{4}(N_cm_c)^2/r_c,
\end{equation}
and the relations
\begin{mathletters}
\begin{eqnarray}
n_c&=&N_c/(4\pi r_c^3/3)\approx 2\times 10^9\,{\rm pc}^{-3}
(N^{-2}_{c,8}v^6_{c,3}m^{-3}_{c,*}),\\
r_c&=&0.2\,{\rm pc}\,(N_{c,8}v^{-2}_{c,3}m_{c,*}),
\end{eqnarray}
\end{mathletters}%
where $N_{c,8}\equiv N_c/10^8, v_{c,3}\equiv v_c/10^3\,{\rm km}\,\,
{\rm s}^{-1}$, and $m_{c,*}\equiv m_c/M_{\odot}$. The central relaxation
time is (Spitzer and Hart (1971)):
\begin{equation}
t_r\approx v_c^3/\biggl[\biggl(\frac{3}{2}\biggr)^{1/2}
4\pi m_c^2n_c\ln(0.4N_c)\biggr]
\approx 0.8\times 10^8\,{\rm yr}\,(N^2_{c,8}v^{-3}_{c,3}m_{c,*}
\Lambda_8^{-1}),
\end{equation}
where $\Lambda_8\equiv \ln(0.4N_c)/\ln(0.4\times 10^8)$. Moreover, the
dynamical time scale $t_d \ll t_r$ is
\begin{equation}
t_d\approx r_c/v_c\approx 200\,{\rm yr}\,(N_{c,8}v^{-3}_{c,3}m_{c,*}),
\end{equation}
and the two-body collision time scale is
\begin{equation}
t_{\rm coll}\approx\frac{1}{n_c\sigma_{\rm coll}v_{\infty}}
\approx 6\times 10^{13}\,{\rm yr}\,(N^2_{c,8}v^{-5}_{c,3}m_{c,*}),
\end{equation}
where $v_{\infty}\approx 2^{1/2}v_c$ is the asymptotic relative velocity,                                      
and $\sigma_{\rm coll}$ is the collision cross section.

The physical source of the nonlinear contributions in the transport equations for the order
parameters $E_c$ and $N_c$ could be energy pumped into the core from binary gravitational
radiation, heat dissipative effects caused e.g., by tidal interactions that can transfer large
amounts of energy to the core in its final stage of evolution, or post-Newtonian contributions in
the self-gravitating interactions of the point masses in the core.

\section {Approximate Solution for the Redshift}

By using the result for the redshift (Cohn 1980; Stuart and Teukolsky 1985):
\begin{equation}
z_c\approx \phi_c\approx 14v_c^2/3\approx 5\times 10^{-5}v^2_{c,3},
\end{equation}
where $\phi_c$ is the central potential, we can replace Eq.(\ref{energy}) by
\begin{equation}
\label{redshift}
{\dot z}_c=(a_1/t_r)z_c-(b_1/t_r^3)z_c^3+(\beta_2/t_r^3)N_c^2z_c
+(1/t_{\rm coll})z_c,
\end{equation}
where $a_1=\alpha_2-\alpha_1$, $b_1=(3N_cm_c/28)^2\beta_1$ and the total 
core mass $N_cm_c$ is constant.

We solve Eqs.(\ref{redshift}), (\ref{Nequation}) and (\ref{mequation})
subject to the constraint: $\tau \leq H^{-1}=2\times 10^{10}\,{\rm yr}$,
where $\tau$ is the remaining time for the gravothermal catastrophe to
drive secular core collapse to completion ($N_c\rightarrow 0$). A solution 
for $\alpha_1$ and $\alpha_2$ (Cohn 1980; Stuart and Teukolsky 1985), using a
matching to Fokker-Planck calculations for the advanced core collapse in the 
absence of collisions, yielded the result $\alpha_1/\alpha_2=0.701$ ($a_1 > 0$). 
It should be noted that the Fokker-Planck approximations used by Cohn (1980) eventually break
down as the number of stellar remnants in the core decreases and large angle scattering and
binary formation become important, so these caculations may not include nonlinear
cooperative phenomena in the last highly dense stage of core evolution.

We shall assume that the parameter $a_1$ remains positive for the nonlinear
equations, Eqs.(\ref{redshift}) and (\ref{Nequation}), that
$t_{\rm coll}=\infty$, that the parameter $\beta_2\approx 0$ and that in the time
scale of the gravothermal catastrophe evolution 
$t_r\approx {\rm const.}$ Then, Eq.(\ref{redshift}) can be written as
\begin{equation}
{\dot z}_c=az_c-bz_c^3,
\end{equation}
where $a=a_1/t_r$ and $ b=b_1/t_r^3$. This equation has an exact time dependent
solution for $a>0$ and $b > 0$ given by
\begin{equation}
z_c=\frac{a^{1/2}}{\{b+Ca\exp[-2a(t-t^\prime)]\}^{1/2}},
\end{equation}
where $C$ is an integration constant. For $b=0$ we get 
\begin{equation}
z_c=C^{-1/2}\exp[a(t-t^\prime)],
\end{equation}
and $z_c\rightarrow\infty$ as $t\rightarrow\infty$, which leads to the 
catastrophic collapse to a black hole. However, for
$b > 0$ we get as $t\rightarrow\infty$:
\begin{equation}
z_c\approx (a/b)^{1/2},
\end{equation}
and if $(a/b)^{1/2} < 0.5$, then the dark cluster can be relativistically stable if 
the adiabatic index $\Gamma_1\approx  5/3$ and the cluster has a Newtonian 
mantle and halo with a mantle radius, $r_m\geq 0.1$ pc (Moffat 1997);
the massive core will contain most of the mass of the galaxy.

\section {Discussion}

Observations of nuclear regions in the centers of galaxies have
detected high velocities of revolution, and luminosity spikes suggesting 
masses between $10^7$ and several $10^9\,M_{\odot}$ (Ford et al., 1994; Miyoshi et al., 1995;
Kormendy and Richstone, 1995; Kormendy et al., 1997). However,  masses for more than 60
other nearby galaxies are of order $\sim10^6\,M_{\odot}$ (Kormendy 1987). For our Galactic
center, the
mass may be significantly less than $10^6\,M_{\odot}$ (Allen and Sanders
1986; Kundt 1990; Sanders 1992). The black hole models of clusters
have been used to explain AGNs, and other intense extragalactic radio
sources powered by jets of plasma and magnetic fields produced by the 
compact massive black hole at the center of the galaxy (Begelman,
Blanford, and Rees 1984). However, Kundt (1996) has formulated an alternative model to
explain AGN observations, based on an application of standard accretion disk theory to the
central parts of the galactic disk. A burning disk can possibly explain the phenomena commonly
attributed to a supermassive black hole. It is not clear at present whether these
models can successfully describe the complex phenomena of AGNs for all the observed galactic
nuclei.

We have demonstrated that by using nonlinear transport equations for the order parameters
$E_c$ and $N_c$, which take into account nonlinear cooperative effects of the subsystem of
point masses, the inner core of a cluster may become relativistically stable, after having been
formed by an epoch of gravothermal catastrophy. Treated as a fluid, the dense core is far from
thermodynamic equilibrium and from known physical processes, we can expect cooperative
phenomena to be important. The nonlinear cooperative 
contributions to the differential equation for the redshift can prevent it
from achieving the critical value, $z_c\sim z_{\rm crit}\sim 0.5$. In the
linearized approximation to the transport equations, the standard result follows
that as gravothermal catastrophe develops, the density of the core increases and
the number of core stellar remnants decreases until $z_c >0.5$, at which
point the core inevitably undergoes gravitational collapse to a black hole.

We have used a heuristic model of the nonlinear cooperative effects in a highly dense core
far from thermodynamic equilibrium, in order to obtain a picture of the 
qualitative behavior of the last epoch of core evolution. A more complete
solution would require a numerical analysis of the problem.
Equations such as Eq.(\ref{energy}) should include stochastic
contributions, leading to non-zero correlation functions between 
two-particle interactions, and use should be made of a Fokker-Planck equation and a
related probability density distribution to analyze the dynamical evolution
of the relativistic core. Moreover, it is unrealistic to ignore collisions, binary
formation and gravitational radiation (Quinlan and Shapiro 1989) in the last stage of
core evolution.

In spite of the shortcomings of our qualitative picture of the last stage of core evolution, it
is clear that non-linear self-organizing, cooperative effects, associated with phase transition
phenomena for
the highly dense cluster cores, can dramatically change the collapse scenarios and possibly
prevent the formation of supermassive black holes.

\acknowledgments
                                                          
I thank the Natural Sciences and Engineering Research Council of
Canada for the support of this work. I thank G. D. Quinlan,  M. A. Clayton, J. L\'egar\'e 
and P. Savaria for helpful discussions.
\vskip 0.2 true in
\centerline{REFERENCES}
\vskip 0.3 true in
\noindent Allen, D. A. and Sanders, R. H. 1986, Nature 319, 191

\noindent Bak, P., Tang, C., and Wiesenfeld, K. 1988, Phys. Rev. A 38, 364

\noindent Bak, P. 1996, How Nature Works, Springer Verlag New York 

\noindent Begelman, M. C., Blanford, R. D., and Rees, M. J. 1984, Rev. Mod. Phys. 56, 255

\noindent Bisnovatyi-Kogan, G. S. and Thorne, K. S. 1970, ApJ, 160, 875

\noindent Bisnovatyi-Kogan, G. S., Merafina, M., Ruffini, R., and Vesperini, E. 1993, 

\noindent ApJ, 414, 187

\noindent Cohn, H. 1980, ApJ, 242, 765

\noindent Colgate, S. A. 1967, ApJ, 150, 163

\noindent Fackerell, E. D. 1970, ApJ, 160, 859

\noindent Fackerell, E. D., 1969, Ipser. J. R. and Thorne, K. S. Comments Astrophys. and Space 

\noindent Phys., 1, 140 

\noindent Ford, H. C. 1994, ApJ, 435, L27

\noindent Goodman, J. and Lee, H. M. 1989, ApJ, 337, 84

\noindent Gold, T., Axford, W. I., and Ray, E. C. 1965, in Quasi-stellar Sources and

\noindent Gravitational Collapse, ed. I. Robinson et al. (Chicago: University of Chicago 

\noindent Press), p. 93

\noindent Haken, H. 1975, Rev. Mod. Phys. 47, 67

\noindent Haken, H. 1983, Synergetics, Springer-Verlag New York.

\noindent Ipser, J. R. 1969, ApJ, 158, 17

\noindent ------. 1970, Ap. and Space Sci. 7, 361

\noindent ------.  1980, ApJ, 238, 1101

\noindent Ipser, J. R., and Thorne, K. S. 1968, Ap J, 154, 251

\noindent Kormendy, J. 1987, Structure and Dynamics of Elliptical Galxies, de Zeeuw, T. (ed.), 

\noindent IAU Symp. 127, Reidel,  p.17
 
\noindent Kormendy, J., and Richstone, D. 1995, ARA\&A, 33, 581

\noindent Kormendy, J., et al. 1997, to be published in ApJ.

\noindent Kundt, W. 1990, Astrophys. and Space Sci. 172, 109

\noindent Kundt, W. 1996, Astrophys. and Space Sci. 235, 319

\noindent Lee, H. M. 1995, MNRAS, 272, 605                                                          

\noindent Lightman, A. P. and Shapiro, S. L. 1978, Rev. Mod. Phys. 50, 437

\noindent Lynden-Bell, D. and Wood, R. 1968, MNRAS, 138, 495

\noindent Lugger, P. M. et al. 1992, Astron. J. 104, 83                           

\noindent Moffat, J. W. 1997 UTPT-97-07. astro-ph/9704232                                

\noindent Maoz, E. 1995, ApJ, 447, L91

\noindent Miyoshi, M., et al. 1995, Nature, 373, 127

\noindent Quinlan, G. D. 1996, New Astron. 1, 255

\noindent Quinlan, G. D. and Shapiro, S. L. 1989, ApJ, 343, 725

\noindent Rasio, F. A., Shapiro, S. L., Teukolsky, S. A. 1988, ApJ, 344, 146

\noindent ------. 1989, ApJ, 336, L63

\noindent Sanders, R. H. 1970, ApJ, 162, 791

\noindent Sanders, R. H. 1992, Nature 359, 131

\noindent Shapiro, S. L. and Teukolsky, S. A. 1985a, Ap J, 298, 34

\noindent ------. 1985b, ApJ, 298, 58

\noindent ------. 1985c, ApJ, 292, L41

\noindent ------. 1986, ApJ, 307, 575

\noindent Spitzer, L. 1971, in Galactic Nuclei, ed. D. O'Connell (Amsterdam: 

\noindent North Holland), p. 443

\noindent Spitzer, L., and Hart, M. H. 1971, ApJ, 164, 399

\noindent Spitzer, L., and Stone, M. E. 1967, ApJ, 147, 519                                                          

\noindent van der Marel, R. P. et al. 1997, Nature, 385, 610 

\noindent Zel'dovitch, Ya. B., and Poduretz, M. A. 1965, Astr. Zh., 42, 963 (English transl., 

\noindent 1966,  Soviet Astr. - AJ, 9, 742)

 \end{document}